\documentclass[runningheads]{svmult}

\usepackage{makeidx}   % allows index generation
\usepackage{graphicx}  % standard LaTeX graphics tool
\usepackage{subeqnar}  % subnumbers individual equations
\usepackage{multicol}  % used for the two-column index
\usepackage{physprbb}  % modified textarea for proceedings,
\makeindex             % used for the subject index
\newcommand{\aj}{AJ}
\newcommand{\apjl}{ApJ Lett.}
\newcommand{\apj}{ApJ}
\newcommand{\aap}{A\&A}
\newcommand{\mnras}{MNRAS}

\begin{document}
\title*{Tracing the Remnants of Powerful Quasars to Probe the IGM}
\toctitle{Tracing the Remnants of Powerful Quasars to Probe the IGM}
% \protect\newline 
% allows explicit linebreak for the table of content
%
\titlerunning{Tracing the Remnants of Powerful Quasars to Probe the IGM}
% allows abbreviation of title, if the full title is too long
% to fit in the running head
%
\author{Torsten~A.~En{\ss}lin\inst{1}
\and Rashid~A.~Sunyaev\inst{1,2}
\and Marcus~Br\"uggen\inst{1,3}}
\authorrunning{T.~A.~En{\ss}lin, R.~A.~Sunyaev, M.~Br{\"u}ggen}
% if there are more than two authors,
% please abbreviate author list for running head
%
%
\institute{Max-Planck-Institut f\"{u}r Astrophysik,
Karl-Schwarzschild-Str.1, Postfach 1317, 85741 Garching, Germany \and
Space Research Institute (IKI), Profsoyuznaya 84/32, Moscow 117810,
Russia \and Institute of Astronomy, Madingley Road, Cambridge CB3 0HA, United Kingdom}

\maketitle              % typesets the title of the contribution

\begin{abstract}
Powerful quasars and radio galaxies are injecting large amounts of
energy in the form of radio plasma into the inter-galactic medium
(IGM).  Once this nonthermal component of the IGM has radiatively
cooled the remaining radio emission is difficult to detect.  Two
scenarios in which the fossil radio plasma can be detected and thus be
used to probe the IGM are discussed: a) re-illumination of the radio
emission due to the compression in large-scale shock waves, and b)
inverse Compton scattered radiation of the cosmic microwave
background (CMB), cosmic radio background (CRB), and the internal
very low frequency synchrotron emission of the still relativistic low
energy electron population. We present 3-D magneto-hydrodynamical
simulations of scenario a) and compare them to existing
observations.  Finally, we discuss the feasibility of the detection of
process b) with upcoming instruments.
\end{abstract}

\section{Fossil Radio Plasma}

%\nocite{1999dtrp.conf..275E,1993MNRAS.264L..25B,1994MNRAS.270..173C,1998ApJ...496..728H,2000ApJ...534L.135M,2000MNRAS.318L..65F,2001ApJ...547L.107F,fabian2001moriond,McNamara2000Paris,heinz2001,schindler2001}

\begin{figure}[t]
\begin{center}
\includegraphics[width=\textwidth]{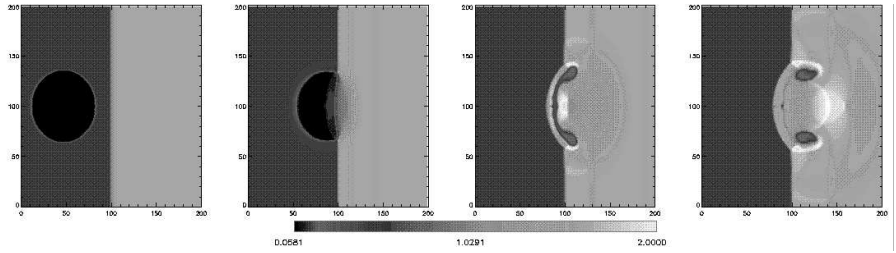}
\end{center}
\caption[]{Shock passage of a hot, magnetized bubble (a radio ghost)
through a shock wave. The flow goes from the left to the right. The
evolution of the mid plane gas density is displayed (white is dense,
black is dilute gas)}
\label{eps1}
\end{figure}

The jets of powerful radio galaxies inflate large cavities in the IGM
that are filled with relativistic particles and magnetic
fields. Synchrotron emission at radio frequencies reveals the presence
of electrons with GeV energies. These electrons have radiative
lifetimes of the order of 100 Myr before their observable radio
emission extinguishes due to radiative energy losses. The remnants of
radio galaxies and quasars are called \lq fossil radio plasma' or a
\lq radio ghosts' \cite{1999dtrp.conf..275E}. Their existence as a
separate component of the IGM is supported by the detections of
cavities in the X-ray emitting galaxy cluster gas \cite[and
others]{1993MNRAS.264L..25B,1994MNRAS.270..173C,1998ApJ...496..728H,2000ApJ...534L.135M,2000MNRAS.318L..65F,2001ApJ...547L.107F,fabian2001moriond,McNamara2000Paris,heinz2001,schindler2001}.
In many cases associated radio emission and in a few cases a lack of
such emission was found, as expected for aging bubbles of radio
plasma. Such bubbles should be very buoyant and therefore rise in the
atmosphere of a galaxy cluster. It is not clear yet if they break into
pieces during their ascent and thereby are slowed down. Another
possibility is that they are able to ascend up to the accretion shock
of a galaxy cluster, where their further rise will be prohibited by
the infalling gas of the accretion onto the cluster.

\section{Shock Wave Re-Illumination}
Whenever a radio ghosts is hit by a shock wave, which may originate
either from a cluster merger or from the steady accretion of gas onto
the still forming large scale structure, it is strongly
compressed. The compression should be adiabatic since typical IGM
shock speeds of a few 1000 km/s are expected to be well below the
internal sound speed of the fossil but still relativistic
plasma. Since the radio plasma has to adapt to the new ambient
pressure, the compression factor can be high and the particles and
magnetic fields can gain a substantial amount of energy. The
synchrotron emission can go up by a large factor, especially at
frequencies which were only a little bit higher than the cutoff
frequency of the uncompressed fossil plasma
\cite{2001A&A...366...26E}. Thus, the radio plasma can be revived to
emit at observing frequencies if it was not too old, a few 100 Myr
inside and few Gyr at the boundary of galaxy clusters.

3-D magneto-hydrodynamical simulations \cite{ensslinbrueggen01} show
that during the traversal of a shock wave, the radio plasma is first
flattened and then breaks up into small filaments, often in form of
one or several tori. The formation of a torus can be seen in
Fig. \ref{eps1}. At places where the hot, under-dense radio plasma
bubble touches the shock wave the balance of pre-shock ram-pressure
and post-shock thermal pressure disappears due to the lack of
substantial mass load of the advected radio plasma. The post shock gas
therefore starts to break through the radio plasma and finally
disrupts it into a torus or a more complicated filamentary structure.

\begin{figure}[t]
\begin{center}
\includegraphics[width=\textwidth]{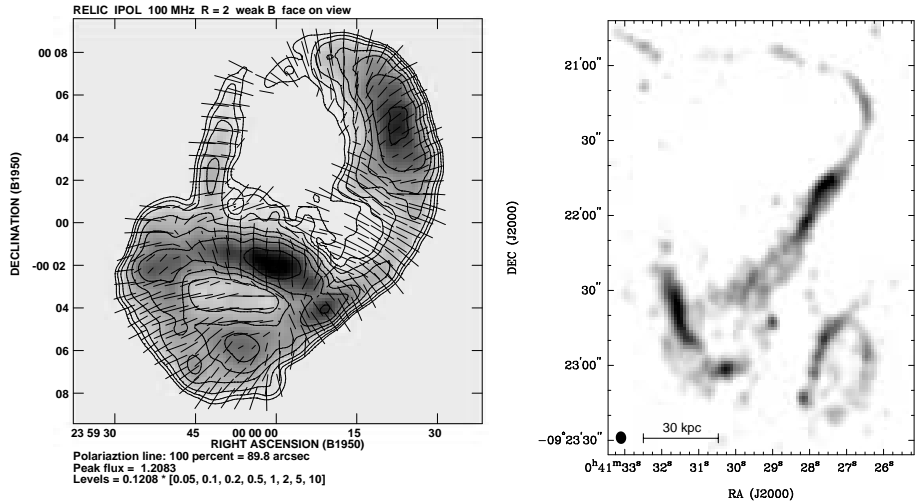}
\end{center}
\caption[]{Left: synthetic radio map of a shocked radio ghost \cite{ensslinbrueggen01}. Right:
observed cluster radio relic in Abell 85 at 1.4 GHz \cite{slee2001}}
\label{eps2}
\end{figure}

The magnetic fields becomes mostly aligned with the filaments
leading to a characteristic polarization signature which can be seen in
the synthetic radio map displayed in Fig. \ref{eps2}.

Polarized radio emitting regions of often filamentary morphologies
could be found in a (recently strongly growing) number of merging
clusters of galaxies. In most cases they are near those places where
shock waves are expected from either observed temperature structures
or comparison of X-ray maps to simulated cluster merger. These radio
sources are called cluster radio relics \cite[and references
therein]{1999dtrp.conf....3F}.  An observed radio map of a filamentary
cluster radio relic in Abell 85 is also displayed in Fig. \ref{eps2}
for comparison. Lower frequency observation show that the upper
filament of this relic forms a closed torus
\cite{2000NewA....5..335G}.

Thus, sensitive observation of cluster radio relics are able to
probe several properties of IGM shock waves
\cite{ensslinbrueggen01}. Since the major diameter of the fossil
radio plasma is approximately conserved, a rough estimate of the
compression factor can be derived from relic radio maps by measuring
the filament diameters. The compression factor depends only on the
pressure jump in the shock and the equation of state of the radio
plasma. Therefore, the shock strength is measurable for a given radio
plasma equation of state. Or, if detailed X-ray maps allow to estimate
the shock strength independently, the equation of state of
radio plasma can be measured.  Furthermore, the total radio polarization of
cluster radio relics (after averaging over the source) contains in
principle enough information to entangle the 3-D orientation
of the shock wave: The sky-projected electric vector is aligned with
the projected shock normal. The polarization fraction is highly
correlated with the angle between the shock normal and the line of
sight \cite{ensslinbrueggen01}.

\section{SSC \& CMB-IC}

\begin{figure}[t]
\begin{center}
\includegraphics[width=\textwidth]{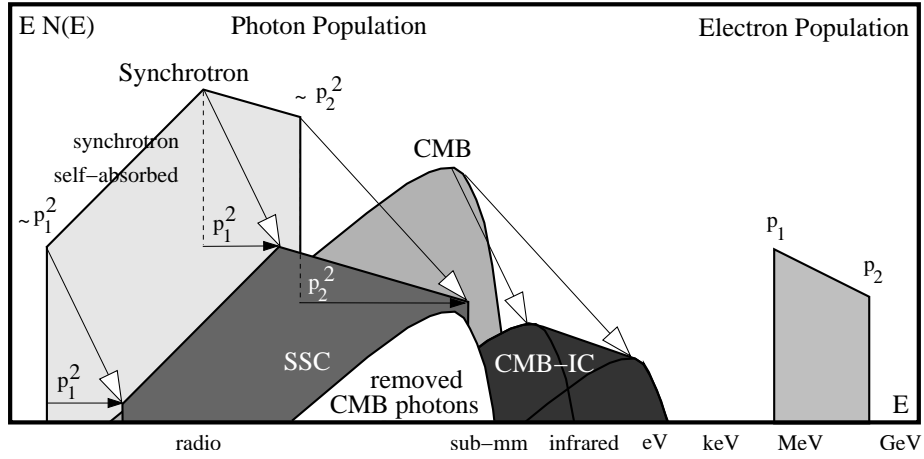}
\end{center}
\caption[]{Sketch of the SSC and the CMB-IC process}
\label{eps3}
\end{figure}

But even very old radio plasma may be detectable by its long lasting
very low frequency radio emission (kHz -- MHz). Even if this emission
is undetectable directly for terrestrial telescopes, it can be
measured indirectly due to the unavoidable inverse Compton (IC)
scattering of the synchrotron photons by their source electrons
\cite{EnsslinSunyaevI01}. These synchrotron self-Comptonized (SSC)
photons have much higher energies and can therefore be in observable
wavebands.  The relativistic electron population also up-scatters
every other present photon field. Photons of the cosmic microwave
background (CMB) but also of the cosmic radio background (CRB) are
removed from their original spectral location and shifted to much higher
frequencies by IC encounters with the fossils radio plasma electron
population. Since the frequency shift is large for IC scattering by
ultra-relativistic electrons, the CMB flux is decremented within the
whole typical CMB frequency range \cite{ensslin2000a}.  The spectral
signature of all these processes are sketched in Fig. \ref{eps3}.

\begin{figure}[t]
\begin{center}
\includegraphics[width=\textwidth]{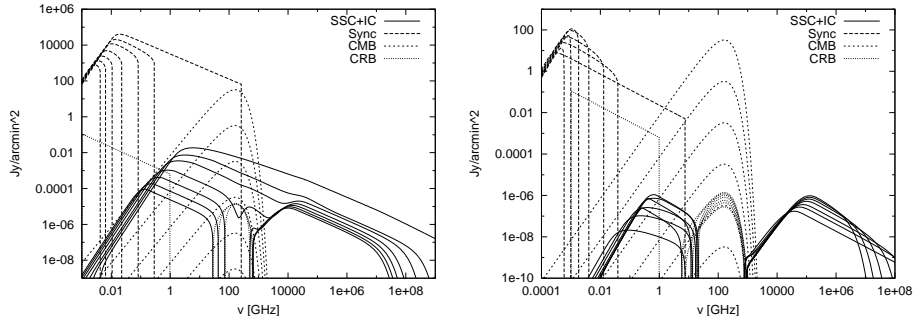}
\end{center}
\caption[]{Central surface brightness of a Cygnus A-like radio cocoon
in a cooling and expanding phase (left) and of a possible radio ghost
produced by our own galaxy during cooling and moderate compression
(right). The synchrotron (long-dashed) and SSC+IC spectra (solid) are
shown for the stages at the jet-power shutdown and for later stages
(left: from top to bottom spectra at ages of 0, 20, 40, 80, 120, 160,
and 200 Myr are displayed; right: from bottom to top spectra at ages
of 0, 1, 2, 4, 6, 8, and 10 Gyr are displayed). In spectral regions,
where the SSC+IC processes lead to a reduction of the brightness below
the CMB brightness, the absolute value of the (negative) SSC+IC
surface brightness is plotted by a dotted line.  The top one of the
short-dashed lines is the CMB spectrum, the short-dashed lines below
this are $10^{-2}, 10^{-4}, 10^{-6}, 10^{-8}, \,\mbox{and}\,10^{-10}$
times the CMB spectrum for comparison of the source to the CMB
brightness. The dotted power-law line at frequencies below 1 GHz is
the cosmic radio background (CRB)}
\label{eps4}
\end{figure}

The strongest sources of such low frequency SSC emission should be
radio lobes of just extinguished powerful radio galaxies (see
Fig. \ref{eps4} and \cite{EnsslinSunyaevI01} for details). If for
example the central engine of Cygnus A would decease today, its
directly observable radio lobe synchrotron emission would vanish
within some 10 Myr due to radiation and adiabatic losses of the
(still) expanding radio plasma. But SSC and CMB-IC emission (or
decrement) can remain for a few 100 Myr. Since the SSC emission is
very sensitive to the compression state of the radio plasma it would
decrease rapidly due to adiabatic expansion during the buoyant rise of
the radio bubble in the cluster atmosphere. The CMB-IC process
is much less sensitive to compression and would start to
dominate the spectrum above 30 GHz after roughly 100 Myr.

Also our own galaxy might have produced radio ghosts during earlier
active phases of the central black hole. Owing to the low density
environment of the local group (compared to the Cygnus A galaxy
cluster) such a ghost is expected to be very relaxed and to have a
very low surface brightness (see Fig. \ref{eps4}). But it may be
detectable because of its large angular scale. If further
environmental compression would have increased its SSC and CMB-IC glow
(in the displayed model a compression by a factor 11 in a 10 Gyr
period was assumed) a future detection may be possible.

The detection of SSC from radio ghosts is an observational
challenge. It would be rewarded by revealing the locations of fossil
radio plasma graveyards. It would provide important information on the
lower end of the relativistic electron population. This would be very
valuable since the electron energy range above a few 10 keV and below
100 MeV is still an unexploited spectral regions. Further, due to the
strong dependence of the SSC emission on the compression stage of
radio plasma, SSC emission is also a sensitive probe of the IGM
pressure.

Due to its broad frequency spectra, it can be probed with several
future high sensitivity instruments, ranging from lowest radio
frequency radio telescopes as GMRT and LOFAR, over microwave
spacecrafts like MAP and PLANCK, balloon and ground based CMB
experiments, and sub-mm/IR projects as ALMA and the HERSCHEL
satellites. A multi-frequency sky survey, as will be provided by the
Planck experiment, should allow to search for the SSC and relativistic
IC spectral signature of many nearby clusters of galaxies and radio
galaxies, at least in a statistical sense by co-adding the signals
from similar sources \cite{ensslin2000a}. In addition to this, there
should be targeted observations of promising candidates, as e.g. the
recently reported X-ray cluster cavities without apparent observable
synchrotron emission. New and upcoming lower frequency ($\le$ 10 GHz)
radio telescopes like LOFAR, GMRT, EVLA, and ATA should have a fairly
good chance to detect such sources. E.g. a Cygnus-A like radio cocoon
should be detectable for these telescopes out to a few 100 Mpc even
$\sim$ 100 Myr after the jetpower shutdown \cite{EnsslinSunyaevI01}.

We hope that this work stimulates observational efforts to exploit the
spectral landscape of the {\it terra incognita} of 1-100 MeV electrons
residing in the intergalactic space via their combined SSC and CMB-IC
emission.

%\begin{thebibliography}{8.}
%\addcontentsline{toc}{section}{References}
%\bibliography{tae}
%\bibliographystyle{plain}
%\bibliographystyle{unsrt}
%\end{thebibliography}

\end{document}